\documentclass[aps,pra,superscriptaddress,twocolumn,showpacs]{revtex4}
\usepackage[colorlinks, linkcolor=blue, citecolor=blue, urlcolor=blue, breaklinks=true]{hyperref}
\usepackage{amsmath}
\usepackage{graphicx}
\usepackage{amssymb}
\usepackage{bbm, color}
\usepackage{relsize}
\usepackage[title]{appendix}
\usepackage{amsthm}
\usepackage{lipsum}
\usepackage{listings}
\usepackage{bbold}

\theoremstyle{plain}
\newtheorem{thm}{Theorem}

\newtheorem{pro}[thm]{Proposition}

\theoremstyle{plain}

\newcommand*{\cl}[1]{{\mathcal{#1}}}
\newcommand*{\bb}[1]{{\mathbb{#1}}}

\newcommand{\proj}[2]{| #1 \rangle\!\langle #2 |}
\newcommand*{\tn}[1]{{\textnormal{#1}}}
\newcommand*{\1}{{\mathbbm{1}}}
\newcommand{\T}{\mbox{$\textnormal{Tr}$}}

\newcommand{\ba}{\begin{eqnarray}}
\newcommand{\ea}{\end{eqnarray}}

\newcommand{\eq}[1]{(\hyperref[eq:#1]{\ref*{eq:#1}})}

\newcommand{\lemm}[1]{\hyperref[lemm:#1]{Lemma~\ref*{lemm:#1}}}

\begin{document}

\title{Upper bounds on the private capacity for bosonic Gaussian channels}
\author{Kabgyun Jeong}
\email{kgjeong6@snu.ac.kr}
\affiliation{Research Institute of Mathematics, Seoul National University, Seoul 08826, Korea}
\affiliation{School of Computational Sciences, Korea Institute for Advanced Study, Seoul 02455, Korea}

\date{\today}

\begin{abstract}
Recently, there have been considerable progresses on the bounds of various quantum channel capacities for bosonic Gaussian channels. Especially, several upper bounds for the classical capacity and the quantum capacity on the bosonic Gaussian channels, via a technique known as quantum entropy power inequality, have been shed light on understanding the mysterious quantum-channel-capacity problems. However, upper bounds for the private capacity on quantum channels are still missing for the study on certain universal upper bounds. Here, we derive upper bounds on the private capacity for bosonic Gaussian channels involving a general Gaussian-noise case through the conditional quantum entropy power inequality.
 \end{abstract} 

\maketitle
\section{introduction}
One of most fundamental and challenging tasks in quantum Shannon theory, a quantum analogue of information theory~\cite{S48}, is to determine the channel capacity of a given quantum channel~\cite{NC00,H16,W17,W18}. In general, the quantum channel transmits a quantum state to another quantum state, and mathematically it is given by a completely positive and trace-preserving map. The channel capacity of a quantum channel is also defined as the maximum rate at which certain (classical, private, or quantum) information can be transmitted reliably through the channel in the limit of vanishing errors. Herein, we restrict to the quantum channel described only by Gaussian unitary transforms over bosonic Gaussian systems with an environmental bosonic Gaussian noise~\cite{HW01,WPG+12,S17}.

The private capacity for a given quantum channel quantifies the ability to transmit a classical private information, and it is maximum rate of the private information in the limit of infinitely many uses of the channel and vanishing errors in the presence of noise through the channel~\cite{L96,D05}. To compute the private capacity, we also need to device the regularization of classical private information, which quantifies the (real) private capacity of the quantum channel. Main difference of the private capacity to the classical capacity is that, in principle, a classical \emph{private} information cannot be accessible from the auxiliary environmental system out of the channel, thus, it is probably applicable for secure quantum communications. Furthermore, it is known that the private capacity is non-additive~\cite{SRS08,LWZG09}, which implies that this quantity is extremely hard to compute.

In this reason, we firstly try to calculate reasonable upper bounds on the private capacity from newly posed quantum entropy power inequality. For the private capacity, previously there have been observed few results on the upper bounds via data-processing inequality and channel simulation~\cite{PLOB17}. See also Refs.~\cite{PBL+18,CM17,RMG18,SWAT18,NAJ19,LTB+19,NPJ20} for related derivations. However, our bounds are more intriguing in the bosonic Gaussian regime of low-energy powers.

Quantum entropy power inequality (\textsc{qEPI}), first proposed by K\"{o}nig and Smith~\cite{KS14}, is the central tool in quantum Shannon theory to estimating the output-entropy of the quantum channel. This inequality states that the output-entropy of a bosonic Gaussian channel, such as a beam-splitter (or amplifier), is always to be increased under two independent input bosonic Gaussian states. Also, the quantum entropy power inequality has been proved several ways with applications~\cite{KS13,KS13+,PMG14,ADO16} and extended to the conditional cases in discrete and Gaussian regimes~\cite{K15,JLJ18,PT18,PH18,HK18}. The power of quantum entropy power inequalities is that those are only carrying the information about von Neumann entropy without details of the quantum state itself. Recently, it has been known that \textsc{qEPI}s have many applications for obtaining upper bounds of the classical capacity~\cite{PH18,JLL19} as well as the quantum capacity~\cite{LLKJ19} on bosonic Gaussian channels with a \emph{general} Gaussian noise one beyond the thermal-noise. The general noise means that it can be possible to take the environmental system as in the form of a squeezed Gaussian (or even non-Gaussian) quantum state~\cite{JLL19}.

In this paper, we consider the conditional quantum entropy power inequality (\textsc{CqEPI}) on the bosonic Gaussian channels, in which the environmental system can be general Gaussian states as an input noise, in order to calculate  upper bounds on the private capacity for those channels. It is not only the first attempt to get a meaningful result on the private capacity using \textsc{CqEPI}, but also gives us an intuition how can we calculate and apply the quantum channel capacity problems on various quantum channels.

This paper is organized as follows. In Section~\ref{pre}, we introduce background notions to understand our results including the quantum entropy power inequalities. We derive universal upper bounds on the private capacity for general bosonic Gaussian channels, and  present a generalized formula in Section~\ref{upper}. In Section~\ref{plot}, we give a specific example for the private capacity with a squeezed thermal noise as one of the general noise model, in order to present physical relevance. Finally, we briefly summarize our results, and comment on a few remarks and open problems in Section~\ref{discussion}.

\section{preliminaries}\label{pre}
\begin{figure}[!t]
 \centering
\includegraphics[width=8.5cm]{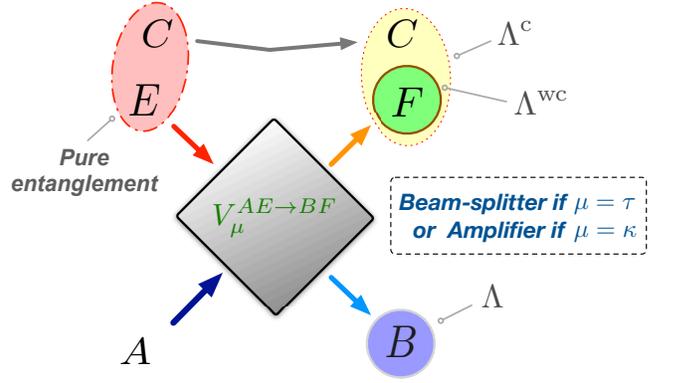}
\caption{A schematic diagram for complementary and weak-complementary channels for the bosonic Gaussian channel $\Lambda$. For a mixed environmental state $\varrho_E$ for the channel $\Lambda$, the purified total system is represented as $\psi_{EC}$.}  
\label{fig1}
\end{figure}

For any Gaussian input state $\varrho_A$ and the environmental system $\varrho_E$, the bosonic Gaussian channel $\Lambda$ via the isometric map can be represented by
\begin{equation}
\Lambda(\varrho_A)=\T_F\left[ V_{AE}(\varrho_A \otimes \varrho_E)V^{\dagger}_{AE}\right],
\end{equation}
where $V_{AE}$ is a symplectic unitary transformation on the Hilbert space $\tn{Sp}(2n_A,\bb{R})\otimes\tn{Sp}(2n_E,\bb{R})$ with the bosonic input mode $n_A$ and the environmental mode $n_E$, respectively. Conversely, its complementary channel $\Lambda^{\tn{c}}$ of the channel $\Lambda$ is naturally defined by
\begin{equation}\label{compure}
\Lambda^\tn{c}(\varrho_A)=\T_B\left[V_{AE}(\varrho_A\otimes\varrho_E)V^{\dagger}_{AE}\right].
\end{equation}
We note that the isometric map $V_{AE}:=V^{AE\to BF}_{\mu}$ in Fig.~\ref{fig1} for the beam-splitting and the amplifying channel has the mixing parameters $\tau\in[0,1]$ and $\kappa\in(1,\infty]$, respectively.

More precisely, we can take two important symplectic unitaries as follows:
\begin{align}
V_\tau&=\exp\left[\arctan\sqrt{\frac{1-\tau}{\tau}}(\hat{a}^\dag\hat{b}-\hat{b}^\dag\hat{a})\right]\;\;\tn{and} \nonumber \\
V_\kappa&=\exp\left[\tn{arctanh} \sqrt{\frac{\kappa-1}{\kappa}}(\hat{a}^\dag\hat{b}^*-\hat{b}^T\hat{a})\right],
\end{align}
where $\hat{a}$ and $\hat{b}$ are annihilation operators of the input and the environment satisfying the canonical commutation relation (CCR), and the superscripts $*$ and $T$ denote the complex conjugate and the transpose operation, respectively. By exploiting CCR algebra, we can also describe the bosonic Gaussian channels in the forms of
\begin{align}
\hat{c}
&=\left\{ \begin{array}{ll} 
\sqrt{\tau}\hat{a}+\sqrt{1-\tau}\hat{b},\ & \tau\in[0,1];\\ & \\
\sqrt{\kappa}\hat{a}+\sqrt{\kappa-1}\hat{b}^\dag,\ & \kappa>1,
\end{array}\right.
\end{align}
where $\hat{c}$ denotes the output bosonic Gaussian state of the channel $\Lambda$.

However, it is important to note that, if the environmental system $\varrho_E$ is a mixed state, we cannot simply obtain the complementary channel uniquely by this technique. Instead, firstly we need to purify the environmental state, and then find the corresponding symplectic unitary operator in the extended Hilbert space. It can be expressed as 
\begin{equation}\label{com}
\Lambda^\tn{c}(\varrho_A)=\T_B\left[V_{AE}\otimes\1_C(\varrho_A\otimes\psi_{EC})(V_{AE} \otimes \1_C)^{\dag}\right],
\end{equation}
where $\psi_{EC}:=\proj{\psi}{\psi}_{EC}$ is a quantum purification satisfying $\T_{C}\psi_{EC}=\rho_E$~\cite{J94,JL16} and the subscript $C$ denotes a reference system. Also, we can define the weak-complementary channel $\Lambda^\tn{wc}$ as the case for which a mixed state $\varrho_E$ is inserted in Eq.~(\ref{compure}), and $\Lambda^\tn{wc}=\Lambda^\tn{c}$ when $\varrho_E$ is any pure state~\cite{CGH06,H07}. In Fig.~\ref{fig1}, we figure out the situation, in which the environment is a mixed state $\varrho_E$.

Now, we review the private capacity on bosonic Gaussian channels, actually, we need two definitions~\cite{WPG+12,L96,D05}. The first one is the `one-shot' private capacity of a quantum channel $\Lambda$, and it is defined by ($\forall\varrho\in\tn{Sp}(2,\bb{R})$ i.e., a single-mode bosonic Gaussian state)
\begin{equation}
P^{(1)}(\Lambda)=\max_{\forall\varrho=\sum_jp_j\rho_j}\left[\chi(\Lambda)-\chi(\Lambda^{\tn{c}})\right],
\end{equation}
where $\chi(\Lambda)=S(\Lambda(\varrho))-\sum_jp_jS(\Lambda(\rho_j))$ is the well-known Holevo capacity~\cite{H98,SW97}, and $S(\varrho)=-\T[\varrho\log\varrho]$ is the von Neumann entropy. The quantity $\chi(\Lambda)-\chi(\Lambda^{\tn{c}})$ is called the classical private information. In bosonic Gaussian regime, it was known that~\cite{PMG14,JLL19} ($\forall\tau\in[0,1]$ and $\forall\kappa\in(1,\infty]$)
\begin{align}
\chi(\Lambda)
&=\left\{ \begin{array}{ll} 
g\left(\tau N+(1-\tau)N_E\right)-g\left((1-\tau)N_E\right)\\ & \\
g\left(\kappa N+(\kappa-1)N_E\right)-g\left(\frac{\kappa-1}{2\kappa-1}N_E\right),
\end{array}\right.
\end{align}
where $N$ and $N_E$ denote the mean photon numbers for the input state and the thermal-noise environment, respectively, and the entropic function $g(x):=(x+1)\log(x+1)-x\log x$.

The second definition is the `regularized' private capacity, which is normally said to be the private capacity.
The (regularized) private capacity of a bosonic Gaussian channel $\Lambda$ under an energy-constraint with the input mean photon number $N$ is given by
\begin{equation}
\cl{P}(\Lambda)=\lim_{n\to\infty}\frac{1}{n}P^{(1)}\left(\Lambda^{\otimes n},\varrho_n \right),
\end{equation}
where $\Lambda^{\otimes n}$ is the independent $n$-copy of the bosonic Gaussian channel and $\varrho_n$ is any input state (i.e., possibly in the entangled state over the channels) in $n$-tensor product of the input Hilbert space. Also we point out that the maximum in the one-shot capacity is taken over all $\varrho_n$ such that $\bar{E}(\varrho_n)\le nN$. If $P^{(1)}(\Lambda)=\cl{P}(\Lambda)$, then we say the private capacity is \emph{additive}, however, generally it is not true~\cite{SRS08,LWZG09}.

The linear relations of \textsc{qEPI}~\cite{KS14,PMG14} are described as follows:
\begin{align} \label{eq:qepi1}
S(\varrho_{X_1}\boxplus_\tau\varrho_{X_2})&\ge\tau S(\varrho_{X_1})+(1-\tau)S(\varrho_{X_2}) \\ \label{eq:qepi2}
S(\varrho_{X_1}\boxplus_\kappa\varrho_{X_2})&\ge\frac{\kappa}{2\kappa-1} S(\varrho_{X_1}) \nonumber\\
&~~+\frac{\kappa-1}{2\kappa-1}S(\varrho_{X_2})+\ln(2\kappa-1),
\end{align}
where $\varrho_{X_1}$ and $\varrho_{X_2}$ are independent input bosonic Gaussian states, and $\boxplus_\tau$ and $\boxplus_\kappa$ represent beam-splitter and amplifier operations with the mixing parameters $\tau\in[0,1]$ and $\kappa \in(1,\infty]$, respectively.

Finally, we introduce two linear versions of conditional quantum entropy power inequality (\textsc{CqEPI}) over two independent input bosonic Gaussian states. For any product states given in the form of $ \varrho_{X_1Z_1}\otimes\varrho_{ X_2Z_2}$, the \textsc{CqEPI}s for bosonic Gaussian channels are expressed as~\cite{K15,PT18}
\begin{align} \label{eq:cqepi1}
S(\varrho_{X_1}\boxplus_\tau\varrho_{X_2}|_{Z_1Z_2})&\ge\tau S(\varrho_{X_1}|_{Z_1})+(1-\tau)S(\varrho_{X_2}|_{Z_2}) \\ \label{eq:cqepi2}
S(\varrho_{X_1}\boxplus_\kappa\varrho_{X_2}|_{Z_1Z_2})&\ge\frac{\kappa}{2\kappa-1}S(\varrho_{X_1}|_{Z_1}) \nonumber\\
&~~+\frac{\kappa-1}{2\kappa-1}S(\rho_{X_2}|_{Z_2})+\ln{(2\kappa-1)},
\end{align}
where the quantum conditional entropy is defined by $S(\rho_{X}|_Z):=S(\rho_{XZ})-S(\rho_Z)$. For convenience, we consider the quantum entropy power inequalities not in the exponential versions but in the linear forms only.

\section{Upper bounds on the private capacity }\label{upper}
Now, we make use of the notation for a bosonic Gaussian channel with the beam-splitter and the amplifier as $\Lambda_{\tau,\rho_E}$ and $\Lambda_{\kappa,\rho_E}$, respectively, where the environmental system $E$ can be any general bosonic Gaussian state with the mean photon number $N_E$. As mentioned above, there exist upper bounds for the private capacity on the thermal-noise Gaussian channels~\cite{PLOB17,PBL+18,NPJ20,RMG18,SWAT18,NAJ19,LTB+19}. Most novel contribution of this work is that, for deriving the upper bound on the private capacity, we exploit the conditional quantum entropy power inequality as well as considering the `general' Gaussian-noise model as the environmental system, while already known results have been used another techniques, for example, data processing inequality.

Suppose that an input bosonic Gaussian state $\rho$ of each channel has the mean photon number $N$ (thus, the multiple input $\rho_n$ has the energy-constraint bounded above by $nN$), then we can describe the upper bound on the regularized private capacity for $\Lambda_{\tau,\rho_E}$ is obtained by
\begin{widetext}
\begin{align} \label{ineq:pcap1}
\cl{P}(\Lambda_{\tau,\rho_E},N)&:=\lim_{n\to\infty}\frac{1}{n}P^{(1)}\left(\Lambda_{\tau,\rho_E}^{\otimes n},\rho_n\right)  \nonumber\\
&=\lim_{n\to\infty}\max_{\bar{E}(\rho_n)\le nN}\frac{1}{n}\left[S\left(\Lambda_{\tau,\rho_E}^{\otimes n}(\rho_n)\right)-\sum_jp_jS\left(\Lambda_{\tau,\rho_E}^{\otimes n}(\rho_j)\right)-S\left(\Lambda_{\tau,\rho_E}^{\tn{c}\otimes n}(\rho_n)\right)+\sum_jp_jS\left(\Lambda_{\tau,\rho_E}^{\tn{c}\otimes n}(\rho_j)\right) \right] \nonumber \\
&\le\lim_{n\to\infty}\max_{\rho_n}\frac{1}{n}\left[S\left(\Lambda_{\tau,\rho_E}^{\otimes n}(\rho_n)\right)+S\left(\Lambda_{\tau,\rho_E}^{\tn{c}\otimes n}(\rho_n)\right)\right]-\lim_{n\to\infty}\min_{\rho_n}\frac{1}{n}\left[S\left(\Lambda_{\tau,\rho_E}^{\otimes n}(\rho_n)\right)+S\left(\Lambda_{\tau,\rho_E}^{\tn{c}\otimes n}(\rho_n)\right)\right]  \nonumber\\ 
&\le\max_{\rho}\left[S\left(\Lambda_{\tau,\rho_E}(\rho)\right)+S\left(\Lambda_{\tau,\rho_E}^{\tn{c}}(\rho)\right)\right]-\lim_{n\to\infty}\min_{\rho_n}\frac{1}{n}\left[S\left(\Lambda_{\tau,\rho_E}^{\otimes n}(\rho_n)\right)+S\left(\Lambda_{\tau,\rho_E}^{\tn{c}\otimes n}(\rho_n)\right)\right] \nonumber\\ 
&=2g(\tau N+(1-\tau)N_E)-\lim_{n\to\infty}\min_{\rho_n}\frac{1}{n}\left[S\left(\Lambda_{\tau,\rho_E}^{\otimes n}(\rho_n)\right)+S\left(\Lambda_{\tau,\rho_E}^{\tn{c}\otimes n}(\rho_n)\right)\right],
\end{align}
\end{widetext}
where the first inequality comes from the minimization of all quantum state $\rho_n$, and the second inequality from the sub-additivity of the von Neumann entropy. We indicate the first term $2g(\tau N+(1-\tau)N_E)$ in the last equality of Eq.~(\ref{ineq:pcap1}) with the term `maximal capacity' of the bosonic Gaussian channel. We know the upper bounds for first two terms follow from the fact that bosonic Gaussian states always fulfill maximal entropies for given first and second moments~\cite{WGC06}: That is, we have 
\begin{align}
\max_{\rho}S(\Lambda_{\tau,\rho_E}(\rho))
&=\max_{\rho}S(\Lambda_{\tau,\rho_E}^{\tn{c}}(\rho)) \nonumber\\
&=g(\tau N +(1-\tau)N_E),
\end{align}
where we take roughly the maximal entropy of the complementary channel $\Lambda_{\tau,\rho_E}^{\tn{c}}$ as equivalent to the maximal entropy of the channel $\Lambda_{\tau,\rho_E}$ in Eq.~(\ref{ineq:pcap1}). 

The main obstacle is how can we calculate the minimal values of the last two terms, which are extremely hard to compute. Actually, this is a challenging topic in quantum Shannon theory, and it has a special name such a `(Gaussian) minimum output entropy conjecture'. In order to obtain a useful bound on the last two terms, we need to use \textsc{qEPI} in Eq.~(\ref{eq:qepi1}) and \textsc{CqEPI} in Eq.~(\ref{eq:cqepi1}) simultaneously. 

Before the detailed proof, we divide the last two terms as follows:
\begin{align}
[\textbf{D.1}]~~\cl{S}_{\tn{MOE}}(\Lambda_\tau)&=\lim_{n\to\infty}\min_{\rho_n}\frac{1}{n}\left[S\left(\Lambda_{\tau,\rho_E}^{\otimes n}(\rho_n)\right)\right] \nonumber\\
[\textbf{D.2}]~~\cl{S}_{\tn{MOE}}(\Lambda_\tau^{\tn{c}})&=\lim_{n\to\infty}\min_{\rho_n}\frac{1}{n}\left[S\left(\Lambda_{\tau,\rho_E}^{\tn{c}\otimes n}(\rho_n)\right)\right],
\end{align}
where it was conjectured that $\cl{S}_{\tn{MOE}}(\Lambda_\tau)=g((1-\tau)N_E)$ for [\textbf{D.1}]~\cite{GGLMS04,GGL+04} and, without loss of generality, we also assume that $\cl{S}_{\tn{MOE}}(\Lambda_\tau^{\tn{c}})=g((1-\tau)N_E)$ roughly for [\textbf{D.2}] case.
However, in our case, the environment and the output of complementary channel are conditioned by the purifying system $C$, while the input and the environmental systems are initially placed on a product state by the definition of the channel. Therefore, we need two-track strategy to bound the private capacity, not only for the channel $\Lambda_{\tau,\rho_E}$ but also for the complementary channel $\Lambda_{\tau,\rho_E}^{\tn{c}}$.

First, we can modify the bound on the definition [\textbf{D.1}] case, that is,
\begin{align} \label{eq:c1-1}
S\left(\Lambda_{\tau,\rho_E}^{\otimes n}(\rho_n)\right)
&\ge\tau S(\rho_n)+(1-\tau)S(\rho_E^{\otimes n}) \nonumber\\  
&=\tau S(\rho_n)+n(1-\tau)S(\rho_E) \nonumber\\ 
&\ge n(1-\tau)S(\rho_E) \nonumber\\
&=n(1-\tau)g(N_E), 
\end{align}
where the first inequality follows from the \textsc{qEPI}, and the first equality comes from independent and identically distributed (i.i.d.) assumption for environmental noise $\rho_E$, and we make use of the non-negativity of the entropy for the second inequality.

Second, the bound on [\textbf{D.2}] can be similarly derived from \textsc{CqEPI}. In this case,  the \textsc{CqEPI} for the bosonic Gaussian complementary channel is given by
\begin{equation}
S(\Lambda^{\tn{wc} \otimes n}_{\tau,\rho_E}(\rho_n)|_C)\ge\tau S(\rho^{\otimes n}_E|_C)+(1-\tau) S(\rho_n|_C),
\end{equation}
however, the input state $\rho_n$ is initially separable from the reference system $C$, and we also take an assumption of independent and identically distributed (i.i.d.) for the environmental noise $\rho_E$.
Then,
\begin{align}\label{ver1}
S(\Lambda^{\tn{wc} \otimes n}_{\tau,\rho_E}(\rho_n)|_C)&:=S(\Lambda^{\tn{c}\otimes n}_{\tau,\rho_E}(\rho_n))-nS(\rho_C) \nonumber \\&
\ge\tau S(\rho^{\otimes n}_E|_C)+(1-\tau) S(\rho_n) \nonumber \\ 
&=(1-\tau) S(\rho_n)-n\tau S(\rho_E) \nonumber \\
&\ge -n\tau g(N_E),
\end{align}
where the first inequality follows from the \textsc{cQEPI}, the second equality comes from i.i.d. assumption for each $\rho_E$'s and $S(\rho_{EC})=0$, so that $S(\rho_E^{\otimes n}|_C)=-S(\rho_C)$, and the last inequality is obtained from the non-negativity of the von Neumann entropy again. Finally, we get the inequality as $S( \Lambda^{\tn{c}\otimes n}_{\tau,\rho_E}(\rho_n))\ge n(1-\tau)S(\rho_E)$ from $S(\rho_E)=S(\rho_C)$. 
Notice that if the environmental system has a Gaussian thermal noise, then $S(\rho_E)=g(N_\tn{th})$, where $N_\tn{th}$ is the mean thermal photon number of the environment, i.e., $\sum_j\frac{\nu_j-1}{2},~\forall j$ for the symplectic eigenvalues $\nu_j$ of a given covariance matrix. For general noise case, $N_E \equiv g^{-1}(S(\rho_E))=N_\tn{th}$. 

Thus, we can conclude that 
\begin{equation} \label{moes}
\cl{S}_{\tn{MOE}}(\Lambda_\tau)+\cl{S}_{\tn{MOE}}(\Lambda_\tau^{\tn{c}})\ge 2(1-\tau)g(N_E),
\end{equation}
where the dependences of $n\to\infty$ and $\rho_n$ (i.e., $\bar{E}(\rho_n)\le nN$) can be dropped. Now, let us combine above Eq.~(\ref{moes}) into Eq.~(\ref{ineq:pcap1}), we can take an upper bound for the private capacity as in the form of
\begin{equation} \label{eq:upper}
\cl{P}(\Lambda_{\tau,\rho_E},N)\le2\left[g(\tau N+(1-\tau)N_E)-(1-\tau)g(N_E)\right].
\end{equation}
We here observe that the upper bound on the private capacity for the beam-splitter channel has double value compare to the upper bound on the quantum capacity case~\cite{LLKJ19}, i.e., $\cl{P}(\Lambda_{\tau,\rho_E},N)\simeq2\cl{Q}(\Lambda_{\tau,\rho_E},N)$. 

Similarly, we can also get the upper bound on the amplifier, which follows from Eq.~(\ref{eq:cqepi2}) above,
\begin{align}
\cl{P}(\Lambda_{\kappa,\rho_E},N)&\le2\Big[g(\kappa N+(\kappa-1)(N_E+1)) \nonumber\\
&~~-\frac{\kappa-1}{2\kappa-1}g(N_E)-\ln(2\kappa-1)\Big].
\end{align}
It is worth to mention that the upper bound increases as the mean photon number $N_E$ of the environment increases. However, it doesn't mean that actual private capacity always depends on the environmental energy.

Along the previous formulation on the classical capacity (see Appendix B in Ref.~\cite{JLL19}), we can introduce a similar argument on the upper bound of the private capacity for bosonic Gaussian channels. Let us $\Gamma_G$ be a single-mode covariance matrix (CvM) for a general bosonic Gaussian noise $\rho_E$ (with the mean photon number $N_E$) satisfying 
\begin{equation}
\det \Gamma_G=(2N_E+1)^2,
\end{equation}
then we have a chance to generalized formula for the strong upper bound on the private capacity.
\begin{pro}
Let $\Lambda_{\tau,\rho_E}$ be a general bosonic Gaussian noise channel with an input bosonic Gaussian state $\rho_A$ with the mean photon number $N$, and the mixing parameter $\tau\in[0,1]$. Then the upper bound on the private capacity of the channel is given by
\begin{align} \label{eq:uppgen}
\cl{P}(\Lambda_{\tau,\rho_E},N)&\le2g(\tau N+(1-\tau)N_E^*) \nonumber\\
&~~-2(1-\tau)g\left(N_E^*\right),
\end{align}
where the general environmental noise $\rho_E$ has a mean photon number $N_E^*:=\tfrac{1}{2}\sqrt{\det \Gamma_G}-1$.
\end{pro}
We notice that the above upper bound, Eq.~(\ref{eq:uppgen}), is weak in the sense of that it could be diverge in the limit of $N\to\infty$, thus it only works in the low-energy limit, however, we take a general Gaussian-noise case $N_E^*$ not on just thermal-noise case with $N_E$.

Now, we briefly mention about the upper bound on the private capacity for the Gaussian amplifier. For the general amplifier channel $\Lambda_{\kappa,\rho_E}$ with $\kappa\in(1,\infty]$, we have~\cite{JLL19}
\begin{align}
\cl{P}(\Lambda_{\kappa,\rho_E},N)&\le2g\left(\kappa N+(\kappa-1)(N_E^*+1)\right) \nonumber\\
&~~-\frac{2\kappa-2}{2\kappa-1}g(N_E^*)-\ln(2\kappa-1)^2.
\end{align}
We can observe that the upper bound on the private capacity has also double value for the upper bound on the classical capacity, $\cl{P}(\Lambda_{\tau,\rho_E},N)\simeq2\cl{C}(\Lambda_{\tau,\rho_E},N)$. Operationally, those three quantities have a relation so that, for any quantum channel $\Lambda$, $\cl{Q}(\Lambda)\le\cl{P}(\Lambda)\le\cl{C}(\Lambda)$, this implies that $\cl{P}(\Lambda)$ can be more tighten the upper bound about 1/2 in bosonic Gaussian channels.

\begin{figure*}[!t]
\centering
\includegraphics[width=13.5cm]{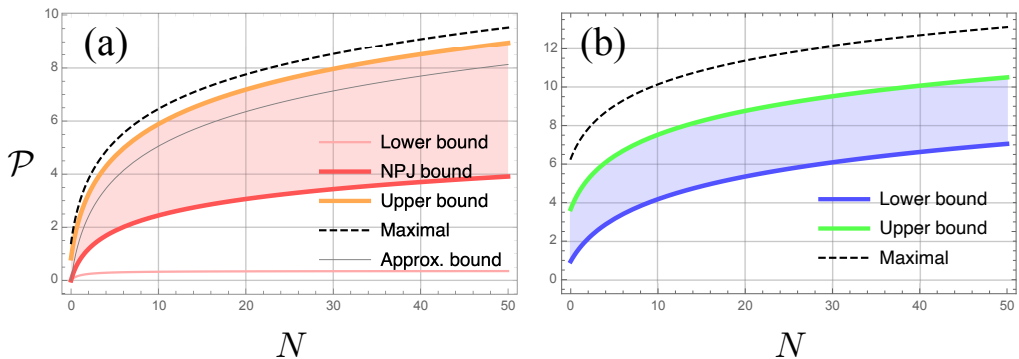}
\caption{Bounds on the private capacity for the bosonic Gaussian channels, when $N$ is the mean photon number of the input state: (a) the beam-splitter case with $\tau=0.85$ and (b) the amplifier with $\kappa=5$. Each upper bounds on private capacities over the beam-splitter and the amplifier (via \textsc{qEPI}s) are described in orange line in (a) and green one in (b), respectively. Specifically in (a), the term `Maximal' means the positive part in Eq.~(\ref{ineq:pcap1}), i.e., $\cl{P}_{\max}(\Lambda_{\tau,\rho_E},N)=\max_\rho\left[S(\Lambda_{\tau,\rho_E}(\rho))+S(\Lambda_{\tau,\rho_E}^{\tn{c}})(\rho)\right]$, and `Lower bound' is taken from Eq.~(\ref{lower}). We denote the enhanced lower bound as `NPJ bound' (with $x=0.95$)~\cite{NPJ20}, and `Approx. bound' means an (estimated) approximate bound on the private capacity as in Eq.~(\ref{approx}).}
\label{fig2}
\end{figure*}

\section{Plots on the bounds for the private capacity }\label{plot}
In the previous section, we have investigated upper bounds on the private capacity for the bosonic Gaussian noise channel in which an environmental system can be any Gaussian-noise state. We should remember that we can take the lower bound from [\textbf{D.1}] and [\textbf{D.2}] from substituting the thermal-noise to the general Gaussian-noise case as
\begin{equation} \label{approx}
\cl{P}(\Lambda_{\tau,\rho_E},N)\ge2\left[g(\tau N+(1-\tau)N_E)-g((1-\tau)N_E)\right],
\end{equation}
since we can fix the lower bound roughly. Similarly, the amplifier's lower bound can be given by
\begin{equation}
\cl{P}(\Lambda_{\kappa,\rho_E},N)\ge2\left[g(\kappa N+(\kappa-1)N_E)-g\left(\tfrac{\kappa-1}{2\kappa-1}N_E\right)\right].
\end{equation}
We plot the upper and lower bounds for the private capacity on the bosonic Gaussian channels with respect to the input state energy $N$ in Fig.~\ref{fig2}. As a special case, it was known that a lower bound for the one-shot private capacity on the thermal-noise channel is given by Eq. (8.9) in~\cite{SWAT18}
\begin{align} \label{lower}
P^{(1)}(\Lambda_{\tau,\rho_E},N)&\ge\cl{P}_L(\Lambda_{\tau,\rho_E},N) \nonumber\\
&\equiv I_\tn{c}(\Lambda_{\tau,\rho_E},N)-I_\tn{c}(\Lambda_{\tau,\rho_E},N^2),
\end{align}
where $I_{\tn{c}}(\Lambda,\varrho):=S(\Lambda(\varrho))-S(\Lambda^c(\varrho))$ denotes the coherent information for the thermal-noise channel for an input $\rho$ with $N$. Also, it was known that the lower bound on the private capacity can be further improved by using a classical-quantum states~\cite{NPJ20}. 

Here we give specific example in order to consider the physical meanings of our results. The non-trivial example on the private capacity of the channel $\Lambda_{\mu,\rho_E}$ is the beam-splitter ($\mu=\tau$) involving the general thermal-noise ($\rho_E=\rho_{\tn{sth}}$), in which the environmental system is the squeezed thermal state $\rho_{\tn{sth}}$ as a general Gaussian noise. In general, the squeezed thermal state has the CvM known in the form of
\begin{equation}
\Gamma_\tn{sth}=(2N_{\tn{th}}+1)
\left( \begin{array}{cc}
e^{-2r} & 0 \\ & \\
0 & e^{2r} \end{array}\right),
\end{equation}
where $N_\tn{th}(=N_E)$ is the mean photon number from the thermal noise, and $r \in [0,\infty)$ is the squeezing parameter. Notice that $\det \Gamma_{\tn{sth}}=(2N_E+1)^2$ is equivalent to as in the general case $\det \Gamma_G$ in Section~\ref{upper}. The squeezed thermal state is the most general single-mode Gaussian state when its mean is placed at the origin, which can be always removed by the local symplectic unitary transformation. Consequently, what we are considering here are general bosonic Gaussian channels, and actually the mean photon number of general $\Gamma_{\tn{sth}}$ is equivalent to the thermal state, i.e., $N_{\tn{sth}}=N_{\tn{th}}=N_E^*$.

\section{discussions}\label{discussion}
Computing the exact channel-capacity on a quantum channel is very hard problem. However, recently it was known that we can effectively obtain the upper bounds through a powerful tool of quantum entropy power inequality. In this study, we have investigated non-trivial upper bounds on the energy-constrained private capacity for bosonic Gaussian channels with mixing parameters $\tau$ and $\kappa$. Here, our principal method is \textsc{CqEPI}, which can be used for obtaining bounds on the output entropy of the complementary bosonic Gaussian channel. Although our results do not allow tighter bounds for the private capacity, those are applicable to more general environmental noise such a squeezed thermal noise, and potentially even at non-Gaussian one. 

Our results showed that the upper bounds for the private capacity have a double value comparing to the classical and the quantum capacity for the bosonic Gaussian channels. In other words, we have some reservations about tight bounds for the three quantities in the operational framework of $\cl{Q}\le\cl{P}\le\cl{C}$. We expect that our work could be extended for the knowledge of the private capacity, which is still far from reaching fully understanding the channel capacity problems in quantum information theory.

\section*{ACKNOWLEDGMENTS}
The author thanks to anonymous valuable comments. This work was supported by Basic Science Research Program through the National Research Foundation of Korea, a grant funded by the Ministry of Education, Korea (NRF-2018R1D1A1B07047512) and the Ministry of Science and ICT (NRF-2020M3E4A1077861).

\end{document}